\documentclass{article}

\usepackage{amssymb}
\usepackage{amsmath}

\def\2{{1\over 2}}

\newcommand{\rf}[1]{(\ref{#1})}
\def\b{\bar}
\newcommand{\ud}{\mathrm{d}}

\newcommand{\p}{\partial}
\newcommand{\bp}{\bar{\partial}}

\newcommand{\mA}{\mathbf{A} }
\newcommand{\mB}{\mathbf{B} }
\newcommand{\mC}{\mathbf{C} }
\newcommand{\mW}{\mathbf{W} }
\newcommand{\mY}{\mathbf{Y} }
\newcommand{\mF}{\mathcal{F} }
\newcommand{\mQ}{\mathcal{Q} }
\newcommand{\m}{\mathfrak{m}}
\newcommand{\di}{\mathbf{div}}
\title{
\bf{Homotopy Lie Superalgebra in Yang-Mills Theory}}
\author{Anton M. Zeitlin\footnote{anton.zeitlin@yale.edu http://pantheon.yale.edu/\~ az84 http://www.ipme.ru/zam.html}   
\footnote{On leave of absence from the St.-Petersburg Division of Steklov Mathematical Institute}\\
Department of Mathematics,\\
Yale University,\\
442 Dunham Lab, 10 Hillhouse Ave\\
New Haven, CT 06511}
\date{}

\begin{document}
\maketitle
\begin{abstract}
The Yang-Mills equations are formulated in the form of generalized Maurer-Cartan equations, such that the corresponding 
algebraic operations are shown to satisfy the defining relations of homotopy Lie superalgebra. 
\end{abstract}
\section{Introduction}
It is well-known in the context of open String Field Theory (SFT) 
(see \cite{siegel}- \cite{taylor1} for review) that in the case of abelian Lie algebra, 
the Yang-Mills equations 
can be obtained from the following relation:
\begin{eqnarray}
Q|\phi\rangle=0, 
\end{eqnarray}
where $Q$ is the BRST operator of open String Theory \cite{brst}, \cite{pol} and 
the state $|\phi\rangle$ corresponds to the operator of ghost number 1:
\begin{eqnarray}
\phi^{(0)}=cA_{\mu}(X)\p X^{\mu}-\p c\p_{\mu}A^{\mu}(X).
\end{eqnarray}
Similarly, the gauge transformation of this abelian gauge field: $A_{\mu}\to A_{\mu}+\p_{\mu}\lambda$ 
can be obtained by means of the transformation
\begin{eqnarray}
|\phi\rangle\to |\phi\rangle +Q|\lambda\rangle,
\end{eqnarray}
where $|\lambda\rangle$ is a state corresponding to the appropriate operator of ghost number 0. 
However, it remains unclear, how to extend this 
cohomological structure to the nonabelian case. In  papers \cite{taylor}, \cite{berkovits},
 the nonabelian versions of Yang-Mills actions were 
obtained from the effective actions of canonical open SFT \cite{witten} and WZW-like superSFT \cite{berksft} correspondingly.
 In this paper, we follow another way: we 
enlarge the space of states associated with the gauge transformations and gauge fields 
by the appropriate states corresponding to the operators 
of ghost number 2 and 3, which altogether form a space of a short chain complex with respect to the operator Q. 
Then, we consider its tensor product 
with some Lie algebra $\mathfrak{g}$ (obviously, this will not spoil the structure of this chain complex, 
since the BRST operator  acts 
trivially on $\mathfrak{g}$). After that, we  explicitly construct the graded antisymmetric bilinear 
and 3-linear operations on this space. We 
show that together with the BRST operator they satisfy the relations of a homotopy Lie superalgebra (this is a 
supersymmetric generalization of \cite{stasheff}, \cite{zwiebach}). 
The bilinear operation appears to be that considered in \cite{zeit3} at the lowest orders in $\alpha'$. 
After these constructions we show that the Yang-Mills equations correspond to the generalized Maurer-Cartan equation 
associated 
with this homotopy algebra, and the associated Maurer-Cartan symmetries correspond 
to the gauge symmetries of Yang-Mills. It is worth 
noting that the equations of motion in 10-dimensional 
supersymmetric Yang-Mills theory was already formulated in the Maurer-Cartan 
form for some differential graded Lie algebra (however, in the different context) in  \cite{movshev}.

\section{Generalized Maurer-Cartan Form of\\ Yang-Mills Eq\-uations.}
{\bf 1. Notation and Conventions.}\\ 
\underline{\bf CFT of open strings and BRST operator.}
We consider the open String Theory in D-dimensional space on the disc conformally mapped to 
the upper half-plane, and we fix the 
operator products between the coordinate fields as follows \cite{pol}: 
\begin{eqnarray}\label{ope}
X^{\mu}(z_1)X^{\nu}(z_2)\sim -\eta^{\mu\nu}\log|z_1-z_2|^2-\eta^{\mu\nu}\log|z_1-\b z_2|^2,
\end{eqnarray}
where $\eta^{\mu\nu}$ is the constant metric in the flat $D$-dimensional space either of Euclidean or Minkowski signature, 
such that the 
mode expansion is
\begin{eqnarray}
&&X^{\mu}(z)=x^{\mu}-i 2p^{\mu}\log|z|^2 +i \sum^{n=+\infty}_{n=-\infty, n\neq 0}\frac{a_n}{n}(z^{-n}+\b z^{-n})\\
&&[x^{\mu},p^{\nu}]=i\eta^{\mu\nu},\quad [a^{\mu}_n, a^{\nu}_m]=\eta^{\mu\nu}n\delta_{n+m,0}.
\end{eqnarray}
We note that we put usual $\alpha'$ parameter equal to 2 \cite{pol}. We also give the expression for the BRST operator 
\cite{brst}, \cite{pol}:
\begin{eqnarray}\label{brst}
Q=\oint dz (cT+bc\p c) , \quad T=-1/2\p X^{\mu}\bp X^{\mu},
\end{eqnarray}
where normal ordering is implicit and $b$, $c$ are the usual ghost fields of conformal weights $2$ and $-1$ correspondingly with the 
operator product 
\begin{eqnarray}
c(z)b(w)\sim \frac{1}{z-w}. 
\end{eqnarray} 
We define the ghost number operator $N_g$ by 
\begin{eqnarray}
N_g=3/2+1/2(c_0b_0-b_0c_0)+\sum^{\infty}_{n=1}(c_{-n}b_n-b_{-n}c_n). 
\end{eqnarray} 
The constant shift (+3/2) is included to make the ghost number of the $SL(2,\mathbb{C})$-invariant vacuum state $|0\rangle$ be 
equal to 0. \\
\underline{\bf {Bilinear operation and Lie brackets.}} In this paper, we  will meet two bilinear operations $[\cdot,\cdot]$, $[\cdot,\cdot]_h$. The first one,
without the subscript,  denotes 
the Lie bracket in the given Lie algebra $\mathfrak{g}$ and the second one,
 with subscript $h$, denotes the graded antisymmetric bilinear operation 
in the homotopy Lie superalgebra. \\
\underline{\bf {Operators acting on differential forms.}} We will use three types of operators acting on differential forms 
(possibly Lie algebra-valued). The first one is the de Rham operator $\ud$. The second one is the {\it Maxwell operator} 
$\mathfrak{m}$, which maps 1-forms to 1-forms. Say, if $\mA$ is 1-form, then 
$\m\mA=(\p_{\mu}\p^{\mu}A_{\nu}-\p_{\nu}\p_{\mu}A^{\mu})dx^{\nu}$, where indices are raised and lowered w.r.t. 
the metric $\eta^{\mu\nu}$. The third operator maps 1-forms to 0-forms, this 
is the operator of divergence $\di$. For a given 1-form $\mA$, $\di\mA=\p_{\mu}A^{\mu}$.

\vspace{3mm}

\noindent{\bf 2. BRST short chain complex.} 
Let's consider the following states
\begin{eqnarray}\label{f0}
&&\rho_{u}=u(x)|0\rangle, \quad \phi_{\mathbf{A}}=(-ic_1A_{\mu}(x)a_{-1}^{\mu}-c_0\p_{\mu}A^{\mu}(x))|0\rangle,\nonumber\\ 
&&\psi_{\mathbf{W}}=-ic_1c_0 W_{\mu}(x)a_{-1}^{\mu}|0\rangle, \quad \chi_{a}=2c_1c_0 c_{-1} a(x)|0\rangle
\end{eqnarray} 
corresponding to the operators
\begin{eqnarray}
u(X), \quad cA_{\mu}(X)\p X^{\mu}-\p c\p_{\mu}A^{\mu}(X),\quad c\p c W_{\mu}(X)\p X^{\mu}, \quad c\p c \p^2 c a(X),
\end{eqnarray}
 associated with functions $u(x)$, $a(x)$ and 1-forms 
$\mathbf{A}=A_{\mu}(x)dx^{\mu}, 
\mathbf{W}=W_{\mu}(x)dx^{\mu}$.  
It is easy to check that the resulting space, spanned by the states like \rf{f0}, is invariant under the action of the BRST 
operator, moreover the following proposition holds.

\vspace{3mm}

\noindent
{\bf Proposition 2.1.} {\it Let the space $\mathcal{F}$ be spanned by all possible states of the form \rf{f0}.  
Then we have a short chain complex:
\begin{eqnarray}\label{complex}
0\to\mathbb{C}\xrightarrow{id}\mathcal{F}^{0}\xrightarrow{Q}\mathcal{F}^{1}
\xrightarrow{Q}\mathcal{F}^{2}\xrightarrow{Q}\mathcal{F}^{3}\to 0,
\end{eqnarray}
where $\mathcal{F}^{i}$ (i=0,1,2,3) is a subspace of  $\mathcal{F}$ corresponding to the ghost number $i$ 
and $Q$ is the BRST operator \rf{brst}.}\\
{\bf Proof.} Really, it is easy to see that we have the following formulas:
\begin{eqnarray}\label{qact}
Q\rho_{u}=2\phi_{\ud u},\quad Q\phi_{\mathbf{A}}=2\psi_{\m \mA},\quad Q\psi_{\mathbf{W}}=-\chi_{\di \mW}, \quad 
Q\chi_{a}=0.
\end{eqnarray}
Then the statement can be easily obtained. $\blacksquare$

\vspace{3mm}

\noindent{\bf Remark.} From \rf{qact}, one can see that the first cohomology module $H^1_{Q}(\mathcal{F})$ can be identified 
with the space of abelian gauge fields, satisfying the Maxwell equations modulo gauge transformations. 

\vspace{3mm}

Now, we introduce the BRST complex which will play the main role in further constructions. Let's consider some Lie algebra $\mathfrak{g}$ 
and take a tensor product of the complex \rf{complex} with $\mathfrak{g}$. In such a way, we get another chain complex: 
\begin{eqnarray}\label{complexb}
0\to\mathfrak{g}\xrightarrow{id}\mathcal{F}_{\mathfrak{g}}^{0}\xrightarrow{\mathcal{Q}}\mathcal{F}_{\mathfrak{g}}^{1}\xrightarrow{
\mathcal{Q}}
\mathcal{F}_{\mathfrak{g}}^{2}\xrightarrow{\mathcal{Q}}\mathcal{F}_{\mathfrak{g}}^{3}\to 0,
\end{eqnarray}
where $\mathcal{F}_{\mathfrak{g}}^{i}=\mathcal{F}^{i}\otimes{\mathfrak{g}}$ and $\mathcal{Q}=Q\otimes 1$. 
In the following, we will keep the same notation \rf{f0} 
for the elements of $\mathcal{F}_{\mathfrak{g}}=\oplus_{i=1}^3\mathcal{F}_{\mathfrak{g}}^{i}$,
 one just need to bear in mind that the 
1-forms and functions, which are associated with the elements of $\mathcal{F}_{\mathfrak{g}}$, are now $\mathfrak{g}$-valued. 

\vspace{3mm}

\noindent{\bf 3. Definition of algebraic operations.} We define  
\begin{eqnarray}
&&[\cdot, \cdot]_h: \mathcal{F}^i_{\mathfrak{g}}\otimes \mathcal{F}^j_{\mathfrak{g}}\to \mathcal{F}^{i+j}_{\mathfrak{g}},\\
&&\label{3lin}[\cdot, \cdot, \cdot]_h: \mathcal{F}^i_{\mathfrak{g}}\otimes \mathcal{F}^j_{\mathfrak{g}}\otimes \mathcal{F}^k_{\mathfrak{g}}\to 
\mathcal{F}^{i+j+k-1}_{\mathfrak{g}},
\end{eqnarray}
which are respectively graded (w.r.t. to the ghost number) antisymmetric bilinear and 3-linear operations (here obviously, 
$\mathcal{F}^i_{\mathfrak{g}}=0$ for $i<0$ and $i>3$). 
The bilinear one is defined by the following relations on the elements of 
$\mathcal{F}_{\mathfrak{g}}$: 
\begin{eqnarray}
&&[\rho_{u},\rho_{v}]_h=2\rho_{[u,v]}, \quad
[\rho_{u},\phi_{\mathbf{A}}]_h=2\phi_{[u,\mathbf{A}]}, \quad 
[\rho_{u},\psi_{\mathbf{W}}]_h=2\phi_{[u,\mathbf{W}]}, \nonumber\\
&&[\rho_u, \chi_a]_h=2\chi_{[\lambda,a]},\quad[\phi_{\mathbf{A}},\phi_{\mathbf{B}}]_h=2\phi_{\{\mathbf{A},\mB\}}, \quad [\phi_{\mathbf{A}},\psi_{\mathbf{W}}]_h= 
-\chi_{\mathbf{A}\cdot\mW},
\end{eqnarray}
where $u$, $v \in \mF^0_{\mathfrak{g}}$, $\phi_{\mathbf{A}}$, $\phi_{\mathbf{B}}\in \mF^1_{\mathfrak{g}}$, 
$\psi_{\mathbf{W}}\in \mF^2_{\mathfrak{g}}$, $\chi_a\in \mF^3_{\mathfrak{g}}$, and we denoted 
\begin{eqnarray}
&&\{\mathbf{A},\mB\}=([A_{\mu},\p^{\mu}B_{\nu}]+[B_{\mu},\p^{\mu}A_{\nu}]+[\p_{\nu}A_{\mu}, B^{\mu}]+ \\
&&[\p_{\nu}B_{\mu}, A^{\mu}]+\p^{\mu}[A_{\mu},B_{\nu}]+\p^{\mu}[B_{\mu},A_{\nu}])dx^{\nu}, \nonumber\\
&&\mA\cdot\mW=[A^{\mu},W_{\mu}]\nonumber.
\end{eqnarray}
The operation \rf{3lin} is defined to be nonzero only when all arguments lie in $\mF^1$ and for $\phi_{\mathbf{A}}$, 
$\phi_{\mathbf{B}}$, $\phi_{\mathbf{C}}$$\in \mF^1$ we have: 
\begin{eqnarray}
[\phi_{\mathbf{A}},\phi_{\mathbf{B}} ,\phi_{\mathbf{C}} ]_h=2\psi_{\{\mA,\mB,\mC\}},
\end{eqnarray}
where we denoted
\begin{eqnarray}
&&\{\mA,\mB,\mC\}=([A_{\mu},[B^{\mu},C_{\nu}]+[B_{\mu},[A^{\mu},C_{\nu}]+[C_{\mu},[B^{\mu},A_{\nu}]+\nonumber\\
&&[B_{\mu},[C^{\mu},A_{\nu}]
+[A_{\mu},[C^{\mu},B_{\nu}]+[C_{\mu},[A^{\mu},B_{\nu}])dx^{\nu}.
\end{eqnarray}
Here, we note that the bilinear operation, defined in this subsection, corresponds to the lowest orders in $\alpha'$ 
of that introduced in \cite{zeit3}. 

\vspace{3mm}

\noindent{\bf 4. Homotopy structure of Yang-Mills theory.} We claim that the graded 
antisymmetric multilinear 
operations, introduced in paragraph 3, satisfy the relations of a homotopic Lie algebra. Namely, the following proposition holds.

\vspace{3mm}

\noindent{\bf Proposition 2.2.} {\it Let $a_1,a_2, a_3, b, c$ $\in$ $\mF$. Then the following relations hold:
\begin{eqnarray}\label{rel}
&&\mQ[a_1,a_2]_h=[\mQ a_1,a_2]_h+(-1)^{n_{a_1}}[a_1,\mQ a_2]_h,\nonumber\\
&&\mQ[a_1,a_2, a_3]_h+[\mQ a_1,a_2, a_3]_h+(-1)^{n_{a_1}}[a_1,\mQ a_2, a_3]_h+\nonumber\\
&&(-1)^{n_{a_1}+n_{a_2}}[ a_1, a_2, \mQ a_3]_h+[a_1,[a_2, a_3]_h]_h-[[a_1,a_2]_h, a_3]_h-\nonumber\\
&&(-1)^{n_{a_1}n_{a_2}}[a_2,[a_1, a_3]_h]_h=0,\nonumber\\
&&[b,[a_1,a_2, a_3]_h]_h=[[b,a_1]_h,a_2, a_3]_h+(-1)^{n_{a_1}n_{b}}[a_1,[b,a_2]_h, a_3]_h+\nonumber\\
&&(-1)^{(n_{a_1}+n_{a_2})n_{b}}[a_1,a_2, [b,a_3]_h]_h.\nonumber\\
&&[[a_1,a_2, a_3]_h,b,c]_h=0.
\end{eqnarray} }
The proof is given in Section 3.

Denoting $d_0=\mQ$, $d_1=[\cdot, \cdot]_h$, $d_2=[\cdot, \cdot, \cdot]_h$, the relations \rf{rel} together 
with condition $\mQ^2=0$ can be summarized 
in the following way: 
\begin{eqnarray}\label{d2}
D^2=0,
\end{eqnarray}
where $D=d_0+\theta d_1+\theta^2d_2$ . Here, $\theta$ is some formal parameter anticommuting with $d_0$ and $d_2$. 
We remind that $d_0$ raises ghost number by $1$, $d_1$ leaves it unchanged while $d_2$ lowers ghost number by 1. Therefore, $d_0,d_2$ are odd elements as well as the parameter $\theta$, but $d_1$ is even. Hence, \rf{d2} gives the following relations:
\begin{eqnarray}
&&d_0^2=0,\quad d_0d_1-d_0d_1=0, \quad d_1d_1+d_0d_2+d_2d_0=0,\nonumber\\ 
&&d_1d_2-d_2d_1=0, \quad d_2d_2=0,
\end{eqnarray}
which are in agreement with \rf{rel}.

\vspace{5mm}

\noindent{\bf Proposition 2.3.} {\it Let $\phi_{\mA}$ be the element of $\mF^1_{\mathfrak{g}}$ associated with 1-form $\mA=A_{\mu}dx^{\mu}$ and 
$\rho_u$ be the element of $\mF^0_{\mathfrak{g}}$ associated with Lie algebra-valued function $u(x)$. 
Then the Yang-Mills equations for $\mA$ and its infinitesimal gauge transformations:
\begin{eqnarray}\label{ym}
&&\p_{\mu}F^{\mu\nu}+[A_{\mu},F^{\mu\nu}]=0, \quad F_{\mu\nu}=\p_{\mu}A_{\nu}-\p_{\nu}A_{\mu}+[A_{\mu},A_{\nu}],\\
&&\label{gt}A_{\mu}\to A_{\mu}+\epsilon(\p_{\mu}u+[A_{\mu},u])
\end{eqnarray}
can be rewritten as follows:
\begin{eqnarray}\label{mc}
&&Q\phi_{\mA}+\frac{1}{2!}[\phi_{\mA},\phi_{\mA}]_h+\frac{1}{3!}[\phi_{\mA},\phi_{\mA},\phi_{\mA}]_h=0,\\
&&\phi_{\mA}\to \phi_{\mA}+\frac{\epsilon}{2}(Q\rho_u +[\phi_{\mA},\rho_u]_h).
\end{eqnarray}
}
{\bf Proof.} Really, from the definition of the brackets, one can see that:
\begin{eqnarray}
&&Q\phi_{\mA}=2\psi_{\mW_1}, \quad W_{1\mu}=\p_{\nu}\p^{\nu}A_{\mu}-\p_{\mu}\p^{\nu}A_{\nu},\nonumber\\
&&[\phi_{\mA},\phi_{\mA}]_h=2\cdot 2!\psi_{\mW_2}, \quad W_{3\mu}=[\p_{\nu}A^{\nu},A_{\mu}]+
2[A^{\nu},\p_{\nu}A_{\mu}]-[A^{\nu},\p_{\mu}A_{\nu}],\nonumber\\
&&\label{gmc}[\phi_{\mA},\phi_{\mA},\phi_{\mA}]_h=2\cdot 3!\psi_{\mW_3},\quad W_{3\mu}=[A_{\nu}, [A^{\nu},A_{\mu}]].
\end{eqnarray}
Summing these identities, we see that equation \rf{mc} is equivalent to 
\begin{eqnarray}
2\psi_{\mW}=0 ,\quad W^{\nu}=\p_{\mu}F^{\mu\nu}+[A_{\mu},F^{\mu\nu}].
\end{eqnarray}
Since we got one-to-one correspondence between the state $\psi_{\mW}$ and 1-form $\mW$, we see that equations \rf{ym} and \rf{mc} 
are equivalent to each other.

Using the formula
\begin{eqnarray}
Q\rho_u +[\phi_{\mA},\rho_u]_h=2\phi_{\ud u+[\mA,u]},
\end{eqnarray}
one obtains that \rf{gmc} coincides with \rf{gt}, which leads to the equivalence of gauge transformations. 
This finishes the proof. $\blacksquare$

\section{Proof of Homotopy Lie Superalgebra Relations}

In this section, we  prove Proposition 2.2. 

Let's start from the first relation:
\begin{eqnarray}\label{der}
\mQ[a_1,a_2]_h=[\mQ a_1,a_2]_h+(-1)^{n_{a_1}}[a_1,\mQ a_2]_h.
\end{eqnarray}
We begin from the case when $a_1=\rho_{u}\in \mF^0_{\mathfrak{g}}$. Then, for $a_2=\rho_v\in \mF^0_{\mathfrak{g}}$ we have:
\begin{eqnarray}
\mQ[\rho_u,\rho_v]_h=4\phi_{\ud [u,v]}=[\rho_u,2\phi_{\ud v}]_h+[2\phi_{\ud u}, \rho_v]_h=
[\mQ\rho_u,\rho_v]_h+[\rho_u,\mQ\rho_v]_h.
\end{eqnarray}
Let  $a_2=\phi_{\mA}\in \mF^1_{\mathfrak{g}}$. Then 
\begin{eqnarray}\label{sum}
\mQ[\rho_u,\phi_{\mA}]_h=4\phi_{\m[u,\mA]}.
\end{eqnarray}
We know that
\begin{eqnarray}
\m[u,\mA]=(\p_{\mu}\p^{\mu}[u,A_{\nu}]-\p_{\nu}\p_{\mu}[u,A^{\mu}])dx^{\nu}.
\end{eqnarray}
At the same time 
\begin{eqnarray}\label{1}
[\mQ \rho_u,\phi_{\mA}]_h=2[\phi_{\ud u},\phi_{\mA}]_h=4\psi_{\mY}, 
\end{eqnarray}
where 
\begin{eqnarray}
&&Y_{\nu}=2[\p_{\mu}u, \p^{\mu}A_{\nu}]+2[A_{\mu},\p^{\mu}\p_{\nu}u]+[\p_{\nu}\p_{\mu}u,A^{\mu}]+\nonumber\\
&&[\p_{\nu}A_{\mu},\p^{\mu}u]+[\p_{\mu}\p^{\mu}u,A_{\nu}]+[\p_{\mu}A^{\mu},\p_{\nu}u]
\end{eqnarray}
and
\begin{eqnarray}\label{2}
[\rho_u,\mQ\phi_{\mA}]_h=4\psi_{[u,\m\mA]}.
\end{eqnarray}
Summing \rf{1} and \rf{2}, we get \rf{sum} and, therefore, the relation \rf{der} also holds in this case. The last 
nontrivial case with $a_1=\rho_u$ is that when $a_2=\psi_{\mW}$. 
We see that 
\begin{eqnarray}
&&\mQ[\rho_u,\psi_{\mW}]_h=-2\psi_{\di [u,\mW]}=-2\psi_{\ud u\cdot W}+2\psi_{[u,\di\mW]}=\nonumber\\
&&[\mQ\rho_u,\psi_{\mW}]_h+[\rho_u,\mQ\psi_{\mW}]_h.
\end{eqnarray}
Let's put $a_1=\phi_{\mA} \in \mF^1_{\mathfrak{g}}$. Then for $a_2=\phi_{\mB} \in \mF^1_{\mathfrak{g}}$, we get 
\begin{eqnarray}
\mQ[\phi_{\mA}, \phi_{\mB}]_h=-2\chi_{\di\{\mA,\mB\}}.
\end{eqnarray}
We find that 
\begin{eqnarray}
&&\di\{\mA,\mB\}=[\p^{\nu}A_{\mu},\p^{\mu}B_{\nu}]+[A_{\mu},\p^{\mu}\p^{\nu}B_{\nu}+[\p^{\nu}B_{\mu},\p^{\mu}A_{\nu}]+\nonumber\\
&&+[\p_{\nu}\p^{\nu}B_{\mu},A^{\mu}]+[\p_{\nu}B_{\mu},\p^{\nu}A^{\mu}]+\p^{\mu}\p^{\nu}([A_{\mu},B_{\nu}]+[B_{\mu},A_{\nu}]=\nonumber\\
&&[\p^{\nu}\p_{\nu}A_{\mu}-\p_{\mu}\p_{\nu}A^{\nu},B^{\mu}]+[\p^{\nu}\p_{\nu}B_{\mu}-\p_{\mu}\p_{\nu}B^{\nu},A^{\mu}]=\nonumber\\
&&(\m \mA)\cdot \mB+(\m \mB)\cdot \mA.
\end{eqnarray}
This leads to the relation:
\begin{eqnarray}
-2\chi_{\di\{\mA,\mB\}}=-2\chi_{(\m \mA)\cdot \mB}-2\chi_{(\m \mB)\cdot \mA}=[Q\phi_{\mA},\phi_{\mB}]-[\phi_{\mA},Q\phi_{\mB}].
\end{eqnarray}
Therefore, \rf{der} holds in this case. 

It is easy to see that relation \rf{der}, for the other values of 
$a_1$ and $a_2$,  reduces to trivial one $0=0$. Thus, we proved \rf{der}.

Let's  switch to the proof of the second relation including the graded antisymmetric 3-linear operation:
\begin{eqnarray}\label{jac}
&&\mQ[a_1,a_2, a_3]_h+[\mQ a_1,a_2, a_3]_h+(-1)^{n_{a_1}}[a_1,\mQ a_2, a_3]_h+\nonumber\\
&&(-1)^{n_{a_1}+n_{a_2}}[ a_1, a_2, \mQ a_3]_h+[a_1,[a_2, a_3]_h]_h-[[a_1,a_2]_h, a_3]_h-\nonumber\\
&&(-1)^{n_{a_1}n_{a_2}}[a_2,[a_1, a_3]_h]_h=0.
\end{eqnarray}
It is easy to see that \rf{jac} is worth proving in the cases, when $a_1\in \mF^0_{\mathfrak{g}}$, $a_2\in \mF^1_{\mathfrak{g}}$, 
$a_3\in \mF^1_{\mathfrak{g}}$ and $a_1\in \mF^1_{\mathfrak{g}}$, $a_2\in \mF^1_{\mathfrak{g}}$, 
$a_3\in \mF^1_{\mathfrak{g}}$. For the other possible values of $a_1,a_2, a_3$, the relation \rf{jac}
 reduces to permutations of the above 
two cases or simple consequences of Jacobi identity for the Lie algebra $\mathfrak{g}$.
 
So, let's consider $a_1=\rho_u\in \mF^0_{\mathfrak{g}}$, $a_2=\phi_{\mA}\in \mF^1_{\mathfrak{g}}$, 
$a_3=\phi_{\mB}\in \mF^1_{\mathfrak{g}}$. In this case, \rf{jac} reduces to
\begin{eqnarray}\label{jac1}
[\mQ \rho_u,\phi_{\mA}, \phi_{\mB}]_h+[\rho_u,[\phi_{\mA}, \phi_{\mB}]_h]_h-[[\rho_u,\phi_{\mA}]_h,\phi_{\mB} ]_h-
[\phi_{\mA},[\rho_u, \phi_{\mB}]_h]_h=0
\end{eqnarray}
or, rewriting it by means of the expressions for appropriate operations, we get:
\begin{eqnarray}
\psi_{\{\ud u,\mA,\mB\}}+\psi_{[u,\{\mA,\mB\}]}-\psi_{\{\mA,[u,\mB]\}}-\psi_{\{\mB,[u,\mA]\}}=0.
\end{eqnarray}
Therefore, to establish \rf{jac1}, one needs to prove:
\begin{eqnarray}
\{\ud u,\mA,\mB\}+[u,\{\mA,\mB\}]-\{\mA,[u,\mB]\}-\{\mB,[u,\mA]\}=0.
\end{eqnarray}
Really,
\begin{eqnarray}
&&\{\mA,[u,\mB]\}=(2[A_{\mu},[\p^{\mu}u,B_{\nu}]]+2[A_{\mu},[u, \p^{\mu}B_{\nu}]]-\nonumber\\
&&2[\p_{\mu}A_{\nu},[u, B^{\mu}]+[\p_{\nu}A_{\mu},[u,B^{\mu}]]+[[\p_{\nu}u,B_{\mu}],A^{\mu}]+\nonumber\\
&&[[u,\p_{\nu}B_{\mu}],A^{\mu}]-[A_{\nu},[\p^{\mu}u,B_{\mu}]]-[A_{\nu},[u,\p_{\mu}B^{\mu}]]+\nonumber\\
&&[\p^{\mu}A_{\mu},[u,B_{\nu}]])dx^{\nu}.
\end{eqnarray}
Rearranging the terms and using Jacobi identity, we find that
\begin{eqnarray}
&&\{\mA,[u,\mB]\}+\{[u,\mB],\mA\}=([u, (2[A_{\mu},\p^{\mu}B_{\nu}]+2[B_{\mu},\p^{\mu}A_{\nu}]+\nonumber\\
&&[\p_{\nu}A_{\mu},B^{\mu}]+[\p_{\nu}B_{\mu},A^{\mu}]-[A_{\nu},\p_{\mu}B^{\mu}]+[\p^{\mu}A_{\mu},B_{\nu}])]+\nonumber\\
&&[A_{\mu},[\p^{\mu}u,B_{\nu}]+[B_{\mu},[\p^{\mu}u,A_{\nu}]+
[A_{\mu},[B^{\mu},\p_{\nu}u]+\nonumber\\
&&[\p^{\mu}u,[B_{\mu},A_{\nu}]+[B_{\mu},[A^{\mu},\p_{\nu}u]]+[\p^{\mu}u, [A_{\mu},B_{\nu}]])dx^{\nu}=\nonumber\\
&&\{\ud u,\mA,\mB\}+[u,\{\mA,\mB\}].
\end{eqnarray}
In such a way we proved \rf{jac1}.

Let's consider the case, when $a_1=\phi_{\mA}\in \mF^1_{\mathfrak{g}}$, $a_2=\phi_{\mB}\in \mF^1_{\mathfrak{g}}$, 
$a_3=\phi_{\mC}\in \mF^1_{\mathfrak{g}}$. For this choice of variables, \rf{jac} has the following form:
\begin{eqnarray}\label{jac2}
&&\mQ[\phi_{\mA},\phi_{\mB}, \phi_{\mC} ]_h+[\phi_{\mA},[\phi_{\mB}, \phi_{\mC}]_h]_h-[[\phi_{\mA},\phi_{\mB}]_h, \phi_{\mC}]_h+\nonumber\\
&&[\phi_{\mB},[\phi_{\mA},\phi_{\mC} ]_h]_h=0
\end{eqnarray}
or, on the level of differential forms,
\begin{eqnarray}\label{formjac}
\di\{\mA,\mB,\mC\}+\mA\cdot \{\mB,\mC\}+ \mC\cdot\{\mA,\mB\}+\mB\cdot \{\mC,\mA\}=0.
\end{eqnarray}
To prove \rf{formjac}, we write the expression for $\mC\cdot\{\mA,\mB\}$:
\begin{eqnarray}\label{abc}
&&\mC\cdot\{\mA,\mB\}=2[C^{\nu},[A_{\mu},\p^{\mu}B_{\nu}]]-2[C^{\nu},[\p_{\mu}A_{\nu},B^{\mu}]+\nonumber\\
&&2[C^{\nu},[\p_{\nu}A^{\mu},B_{\mu}]]+[C^{\nu},[\p_{\nu}B_{\mu},A^{\mu}]]-[C^{\nu},[A_{\nu},\p^{\mu}B_{\mu}]]+\nonumber\\
&&[C^{\nu},[\p^{\mu}A_{\mu},B_{\nu}]]=-([C^{\nu},[\p^{\mu}B_{\nu},A_{\mu}]]+[C^{\nu},[\p_{\mu}A_{\nu},B^{\mu}]]+\nonumber\\
&&[\p^{\mu}B_{\nu},[C^{\nu},A_{\mu}]]+[\p_{\mu}A_{\nu},[C^{\nu},B^{\mu}]]+[C^{\nu},[A_{\nu},\p_{\mu}B^{\mu}]]+\nonumber\\
&&[C^{\nu},[B_{\nu},\p_{\mu}A^{\mu}]])+[A_{\mu},[C^{\nu},\p^{\mu}B_{\nu}]]+[B_{\mu},[C^{\nu},\p_{\mu}A_{\nu}]]-\nonumber\\
&&[C^{\mu},[A_{\nu},\p_{\mu}B^{\nu}]]-[C^{\mu},[B_{\nu},\p_{\mu}A^{\nu}]].
\end{eqnarray}
In order to obtain the last equality, we have used Jacobi identity from $\mathfrak{g}$. Now, we observe that adding to \rf{abc} its 
cyclic permutations, that is $\mA\cdot \{\mB,\mC\}$ and $\mB\cdot \{\mA,\mC\}$, one obtains that the sum of cyclic permutations 
of terms in circle brackets 
(see last equality of \rf{abc}) gives $\di \{\mA,\mB,\mC\}$ while all other terms cancel. This proves  relation \rf{formjac} and, 
therefore, 
\rf{jac2}. Hence we proved \rf{jac}.  

The relations left are:
\begin{eqnarray}
&&[b,[a_1,a_2, a_3]_h]_h=[[b,a_1]_h,a_2, a_3]_h+(-1)^{n_{a_1}n_{b}}[a_1,[b,a_2]_h, a_3]_h+\nonumber\\
&&(-1)^{(n_{a_1}+n_{a_2})n_{b}}[a_1,a_2, [b,a_3]_h]_h,\nonumber\\
&&[[a_1,a_2, a_3]_h,b,c]_h=0.
\end{eqnarray}
However to prove the first one, it is easy to see, that this 
relation is nontrivial only, when $b\in \mF^0_{\mathfrak{g}}$ and $a_i\in\mF^1_{\mathfrak{g}}$. 
Therefore, it becomes a consequence of Jacobi identity from $\mathfrak{g}$. The second one is trivial since the 3-linear operation 
takes values in $\mF^2_{\mathfrak{g}}$ and it is zero for any argument lying in $\mF^2_{\mathfrak{g}}$. 

Thus,  Proposition 2.2. is proved.
\section{Conclusion} 

In this paper, we have shown that the equations of motion of Yang-Mills theory possess a formal Maurer-Cartan formulation. 
We have noted that the bilinear operation in the homotopy Lie 
superalgebra which we considered here, corresponds to the lowest 
orders in $\alpha'$ of that introduced in \cite{zeit3}. One might expect, as we already mentioned in \cite{zeit3}, that 
extending our formalism to all $\alpha'$ corrections, 
we would be able to reproduce the equations of motion corresponding to nonabelian Born-Infeld theory which is the conformal 
invariance condition of the associated sigma model. 

The same approach should be appropriate for gravity: in papers \cite{zeit3}, 
\cite{zeit}, \cite{zeit2} motivated by the structures 
from closed SFT \cite{zwiebach}, we constructed the bilinear operations on 
the corresponding operators which should be associated with some homotopy Lie algebra. In this case, the first order 
formalism, introduced in \cite{lmz}, \cite{zeit2}, looks very 
promising since the associated CFT is the simplest possible and 
the geometric context is undestroyed. Therefore, we are looking forward to introduce the homotopy Lie algebra structure 
in Einstein equations.

\section*{Acknowledgements}

I am grateful to A.S. Losev for 
drawing my attention to this problem. 
I would like to thank I.B. Frenkel, M. Kapranov and G. Zuckerman for numerous discussions on the subject 
and also I.B. Frenkel and N.Yu. Reshetikhin for their permanent encouragement and support.

\end{document}